\documentclass[prb,twocolumn,amsmath,amssymb,floatfix]{revtex4}
\usepackage{graphicx}

\newcommand{\be}{\begin{eqnarray}}
\newcommand{\ee}{\end{eqnarray}}

\newcommand{\bfr}{{\bf r}}
\newcommand{\bfq}{{\bf q}}

\newcommand{\bfk}{{\bf k}}

\newcommand{\bfR}{{\bf R}}

\newcommand{\wbe}{\begin{widetext}}
\newcommand{\wee}{\end{widetext}}
\newcommand{\oncite}{\onlinecite}

\begin{document}

\title{Tuning the Kosterlitz-Thouless transition  to zero temperature \\
in Anisotropic Boson Systems}

\author{Jhih-Shih You$^{1,2}$, Hao Lee$^{1,2}$, Shiang Fang$^{1,2}$, Miguel A. Cazalilla$^{3,4,5}$, Daw-Wei Wang$^{1,2}$}


\affiliation{$^{1}$ Physics Department and Frontier Research Center on Fundamental and Applied Sciences of Matter, National Tsing-Hua University, Hsinchu,
Taiwan
\\
$^{2}$ Physics Division, National Center for Theoretical Sciences,
Hsinchu, Taiwan
\\
$^3$ Centro de Fsica de Materiales CSIC-UPV/EHU. Paseo Manuel de Lardizabal 5, E-20018 San Sebastian, Spain
\\
$^4$ DIPC, Paseo Manuel de Lardizabal 4, E-20018 San Sebastian, Spain
\\
$^5$ Graphene Research Centre National University of Singapore, 6 Science Drive 2, Singapore 117546.
}

\date{\today}

\begin{abstract}
We study the two-dimensional Bose-Hubbard model with anisotropic hopping. Focusing on the effects of anisotropy on the superfluid properties such like the helicity modulus and the normal-to-superfluid (Berezinskii-Kosterlitz-Thouless, BKT) transition temperature,
two  different  approaches are compared: Large-scale Quantum Monte Carlo simulations  and
the self-consistent harmonic approximation (SCHA).  For the latter, two different formulations are considered, one applying near the isotropic limit and the other applying in the extremely anisotropic limit.  Thus we find that  the SCHA provides a reasonable
description of superfluid properties of this system provided the appropriate type of formulation is employed. The accuracy of the SCHA
in the extremely anisotropic limit, where the BKT transition temperature  is tuned to zero
(i.e. into a  Quantum critical point) and  therefore quantum fluctuations play a dominant role, is particularly striking.
\end{abstract}

\maketitle
\section{Introduction}

In recent years, much progress has been made in ultracold  atoms loaded in optical lattices.\cite{Bloch,Zoller,Zwerger} Several experimental groups have demonstrated the large tunability of such systems
by driving them from a superfluid to a Mott  insulator phase (and vice versa) in various  dimensions and lattice
geometries.\cite{Bloch,Esslinger,ChengChin,Troyer,XiboZhang,Becker10,Haller}
By varying the laser intensity along one or several directions,  experimentalists can control
the hopping anisotropy for the atoms in the optical lattice.~\cite{Esslinger,Hadzibabic,ChengChin,Haller}
Thus, some of these experiments have started to explore the fascinating behavior of
ultracold atoms confined to low dimensions.~\cite{Esslinger,Hadzibabic,ChengChin,Haller}
This control makes it also possible to
study dimensional crossovers~\cite{Esslinger, Cazalilla2,Cazalilla3, Iucci,Mathey} as well as a wide
range of  other phenomena,~\cite{Esslinger,Haller} which are also relevant for the understanding of complex
solid-state systems such as layered superconductors\cite{Minnaghen,Shenoy,Olsson,Giamarchi} and
anisotropic magnetic materials.~\cite{Starykh,Kohno,Giamarchi08,Ruegg08}
\begin{figure}[b]
\centering
\includegraphics[width=7cm]{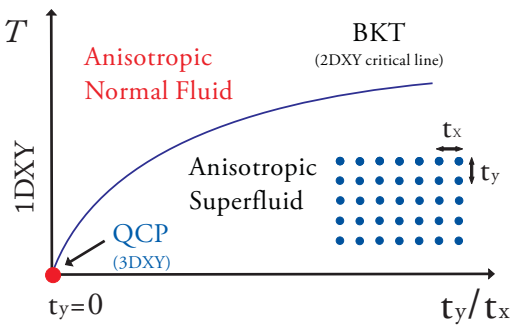}
\caption{Schematic Phase Diagram for the Bose-Hubbad model with hopping anisotropy ratio in two dimensions.}
\label{Tc1}
\end{figure}

Indeed, matter in low dimensions is known to display a wide range exotic properties, which
are otherwise hard to come across in three dimensional systems. These
include  fractionalization of quantum numbers,~\cite{fractional}
critical states lacking long range order,~\cite{Cazalilla_RMP,Esslinger,Haller,Hadzibabic} and
topological phase transitions that cannot be characterized
by an order parameter such as the
Berezinskii-Kosterlitz-Thouless (BKT) transition.~\cite{Kosterlitz,Bishop,Hadzibabic}
The question of how these exotic properties evolve as  low-dimensional
systems are coupled and become, by virtue of the coupling,  higher dimensional systems
has attracted a great deal of experimental and theoretical attention
in recent years.~\cite{Giamarchi08,Ruegg08,Esslinger,Cazalilla2,Cazalilla3,Iucci,Mathey,Cazalilla_RMP,Shlyapnikov,Kastberg,Freericks}

%
%

 In bosonic systems, such as ultracold gases of bosonic atoms or  molecules, as
 well as in  anisotropic magnetic materials,~\cite{Cazalilla_RMP}
 a theoretical analysis of the dimensional crossovers and
 other interesting phenomena such as deconfinement transitions~\cite{Cazalilla2,Cazalilla3}
 can be carried out through a combination of perturbative
 renormalization-group (RG) and mean-field theory (MFT) approaches.
 MFT assumes the existence of a Bose-Einstein condensate
and it is expected to be  a reliable description of the anisotropic superfluid phase only if the
 crossover takes place from one  to three
 dimensions.  Howecer, when trying to describe the crossover
  from one (1D) to two dimensions (2D) at finite temperatures,
MFT breaks down because  bosons in two dimensions fail to  condense
at all temperatures except at $T=0$. Nevertheless, under such conditions
a qualitative understanding of the properties
 of the anisotropic superfluid phase can be still obtained by means of perturbative
 RG and variational methods, as we shall demonstrate below.
 However, an independent check of these approximated methods is still required.

It is worth noting that
the superfluid properties appear to be strongly dependent on the system dimensionality.~\cite{Bishop,Taniguchi10,Nelson,GiamarchiShastry,Affleck,Cazalilla4} Experimentally, a
superfluid response has been observed at finite temperatures  in both two-\cite{Bishop} and
one-dimensional~\cite{Taniguchi10} interacting boson systems, which
lack of a Bose-Einstein condensate. However, the origin of  the superfluidity
in these two cases is very different:~\cite{Nelson,Cazalilla4}
Whereas in 2D the superfluid response is essentially a thermodynamic
phenomenon that is quantified by the helicity modulus,~\cite{Nelson}
in 1D it is a dynamic property as the helicity modulus
vanishes at all temperatures  (the helicity modulus at zero temperature is obtained by
taking the $T\to 0$ limit \emph{after} taking the thermodynamic limit~\cite{GiamarchiShastry,Cazalilla4}).
The vanishing helicity modulus  of the 1D Bose fluid is in stark
contrast with the universal jump exhibited by the helicity modulus across  the BKT transition.
As it will be discussed below,  making the hopping amplitude in one direction
 vanishingly small,  drives the BKT transition temperature to the absolute zero (at $T=0$) and
the transition thus becomes a 3DXY quantum critical point (QCP) at the end of a line of classical 2DXY critical points (cf. Fig~\ref{Tc1}).
 This critical line  separates the anisotropic normal and superfluid phases. Therefore,
it can be argued that the helicity modulus vanishes in 1D Bose fluids because these fluids
share the same superfluid properties as the  \emph{normal} fluid phase in the limit of vanishing anisotropy ratio.
This can be seen by noticing the the helicity modulus of the 1D Bose fluid can be obtained
by approaching the 1D limit  along a finite temperature (i.e. $T > 0$) trajectory (cf. Fig.~\ref{Tc1}), which means
no critical line is crossed for sufficiently small starting anisotropy ratio.  On the other hand,
approaching the 1D limit along the $T=0$ line necessarily implies reaching the QCP first,
which is a thermodynamic singularity (cf. Fig~\ref{Tc1}).

 Indeed, the variety of phenomena that can be studied
 in anisotropic bosonic systems is very wide.~\cite{Cazalilla2,Cazalilla3,Cazalilla_RMP}
 In this work, we focus on understanding the properties of the anisotropic
 superfluid phase that can be realized in e.g. two-dimensional optical lattices with hopping anisotropy.
 However, our results can be also of relevance to much more complex solid-state systems, such like
anisotropic magnetic insulators.~\cite{Giamarchi08,Ruegg08,Cazalilla_RMP}
 In particular, we are interested in understanding how the hopping anisotropy
affects  the  properties of the superfluid phase (i.e. the helicity modulus) and the Berezinskii-Kosterlitz-Thouless (BKT)
transition temperature from the superfluid  to the normal fluid phase.
As the BKT transition temperature is tuned towards $T=0$ by the hopping
anisotropy,   the importance of quantum fluctuations is enhanced. Thus, we expect
that this will lead to important renormalization effects on the parameters of the effective
2DXY model that describes the line of classical critical points. Below we shall
rely on the self-consistent harmonic approximation (SCHA) to various limits of the quantum rotor
model to estimate such renormalization effects.  The results of the calculations based on the SCHA
for the critical temperature and the helicity modulus will compared with Quantum Monte
Carlo simulations.

\begin{figure}[t]
\centering
\includegraphics[width=6cm]{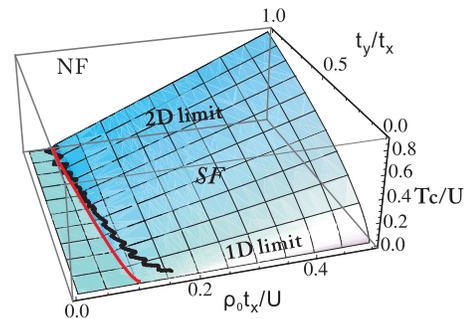}
\caption{The complete phase diagram of an anisotropic Hubbard model as a function of anisotropy ratio, $t_y/t_x$, interaction strength, $\rho_0 t_{\alpha}/U$, and temperature, $T/U$. Note that we keep $U$ to be a constant and vary other quantities for convenience. This is calculated by SCHA as described in the text. The red line guides zero temperature transitions (from superfluid to Mott insulator phase), while the black line is calculated by QMC at very low temperature ($T/U=0.01$) for comparison.  Here we set chemical potential $\mu/U=0.375$(the mean boson occupation $\rho_0\sim1)$ to get the QMC results.}
\label{phase}
\end{figure}

 Of course,  the effect of  quantum fluctuations is enhanced not only by the anisotropy but also by the inter-particle interaction, which
can  drive a quantum phase transition from the superfluid  phase to a Mott insulator phase at integer fillings. In various dimensions,
such superfluid-to-Mott insulator transition has been extensively studied both experimentally and theoretically, mainly in isotropic systems \cite{Bloch,Esslinger,ChengChin,Troyer,XiboZhang,Becker10,Haller}. However,  the combined effect of anisotropy and
inter-particle interactions in enhancing the quantum fluctuations and destroying superfluidity in two dimensional Bose systems
has not been studied so far.  In this paper, we shall show how both quantum and thermal fluctuations can be treated on equal footing in the study of anisotropic superfluid in a 2D optical lattice. Our results can be summarized in the phase diagram shown in Fig. \ref{phase}.

The outline of this article is as follows: In section~\ref{sec:models} we introduce the relevant
low energy models that we use to describe the anisotropic Bose-Hubbard model (BHM) in 2D. Several analytic and numerical
approximations to the BHM are also discussed there.  We also discuss the problem of how to estimate the
``phase-stiffness'' parameters of  the 2DXY model that describes the critical line of BTK transitions
separating the normal and superfluid phases at finite temperature (cf. Figs.~\ref{Tc1},\ref{phase}).
 In section~\ref{sec:small_anisotropy},  the effects of  thermal fluctuations and interaction on helicity modulus  from small to intermediate anisotropy are discussed.  In section~\ref{sec:large_anisotropy} we explore these renormalization effects in the extremely anisotropic regime,  as well as the  behavior of the BKT critical temperature.   The conclusions of this work can be found in section~\ref{sec:conclusion}. The appendixes contain the technical details of SCHA  calculations in the various limits of the quantum rotor model.

\section{Models and Methods}\label{sec:models}

The anisotropy of the single particle tunneling
can be easily realized in an  optical lattice by using different laser intensities for the standing waves
in the $x$ and $y$ directions. In the limit of a deep lattice, in which essentially all particles reside
in the lowest Bloch band,  the system can be described by the single-band Bose Hubbard model:
\be
{H}=-\sum_{\langle i,j\rangle} t_{ij}\:  {b}^{\dagger}_i {b}_j -\mu\sum_i \hat{n}_i, +\frac{U}{2}\sum_i {n}_i ({n}_i-1)\label{eq:bhm}
\ee
where ${b}^\dagger_i$ is bosonic creation operator on site $\bfR_i$, and ${n}_i={b}^{\dagger}_i {b}_i$ is boson ocupation operator. The tunneling amplitude is $t_{ij}=t_{x}$ ($t_{ij}= t_y$), if $\bfR_i-\bfR_j=\pm{a}\hat{\mathbf{x}}$ ($\bfR_i-\bfR_j=\pm{a}\hat{\mathbf{y}}$) with
$a$ being the lattice constant. $U$ and $\mu$ are the on-site interaction and chemical potential, respectively.

\subsection{The XY Model and the Self-consistent Harmonic Approximation (SCHA)}\label{sec:xy}

 Deep into the superfluid phase, the low-temperature behavior of the system is largely dominated by phase fluctuations. For a sufficiently large values of $U$ and for the \emph{bare} anisotropy ratio parameter $\eta_0 = J^0_y/J^0_x\approx 1$,
we can represent  the boson operator as $b_i=\sqrt{\rho_0+\delta\rho_i}\, e^{\textmd{i}\theta_i}$, where $ \rho_0$ is the mean boson occupation, while $\theta_i$ and $\delta\rho_i (\ll \rho_0)$ describe the phase and density fluctuations at site $\mathbf{R}_i$, respectively. After  integrating out the density fluctuations, the partition function ($Z$) of the system can be written as a (Feynman) functional integral,  $Z=\int {\cal D}\theta\,   e^{-S_{\mathrm XY}[\theta]}$, where ($\hbar = 1$)
\be
S_{\mathrm{XY}}[\theta]= \int^{\beta}_0d\tau \left[\sum_i\frac{(\partial_{\tau}\theta_i)^2}{2U}-
\sum_{\langle i,j\rangle}2J^0_{ij}\cos(\theta_i-\theta_j)\right] \label{eq:qrotor}
\label{eq:sxy}
\ee
is the two-dimensional $O(2)$ quantum rotor model, with  $J^0_{ij}=\rho_0 t_{ij}$  being the \emph{bare} Josephson coupling
and $\beta=1/T$  the inverse  of the absolute temperature in units where the Boltzmann constant  $k_B = 1$.

 At sufficiently high temperatures, the imaginary time ($\tau$) dependence of the phase $\theta_i$ can be neglected and the model
 in Eq.~\eqref{eq:sxy} becomes the classical ferromagnetic $XY$ model:
\begin{equation}
S_{\mathrm{C-XY}}[\theta]= -2\beta
\sum_{\langle i,j\rangle} J^0_{ij}\cos(\theta_i-\theta_j).\label{eq:cxy}
\end{equation}
In  2D this model has two distinct phases:
in the high temperature regime, the orientation of the rotors described by the phase $\theta_i$ is disordered. The phase correlations are short ranged, i.e. $g_{ij} = \langle e^{i\theta_i} e^{-i \theta_j}\rangle\simeq e^{-|\mathbf{R}_i - \mathbf{R}_j|/\xi(T)}$, where the correlation length $\xi(T) \gg a$. Such behavior
corresponds to a normal phase. On the other hand, in the low temperature regime, the phase correlations
decay algebraically, i.e. $g_{ij} \simeq |\mathbf{R}_i - \mathbf{R}_j|^{-\alpha(T)}$, where the exponent
$\alpha(T)$ is finite and related to the thermodynamic phase stiffness. This
behavior corresponds to a superfluid phase exhibiting quasi-long range order.  The latter implies
the absence of a Bose-Einstein condensate,  but the finite phase stiffness ($J^0_{ij}$) means that the system can sustain
superflows at all  temperatures below the Berezinskii-Kosterlitz-Thouless (BKT) temperature, $T_{c}$. Above such
temperature,  vortices and anti-vortices unbind and destroy the superfluid properties
of the system. The vortices (anti-vortices) are singular configurations of the phase $\theta_i$, where the latter  winds out by positive (negative) integer multiples of $2\pi$ around a discrete set of points on the plane.

 The picture described above relies on the classical (high temperature) limit of the quantum rotor model,
where the first term in Eq.~\eqref{eq:sxy} ($\propto \left(\partial_{\tau} \theta_i \right)^2$)) is neglected. In other words,
if we expand
\begin{equation}
\theta_i(\tau) =  \frac{1}{\beta}\sum_{\omega_n} e^{-i\omega_n \tau} \: \theta_i(\omega_n),
\end{equation}
where $\omega_n = \frac{2\pi}{\beta} n$ ($n$ being an integer),  the high temperature limit only takes into account the fluctuations of the $\theta_i(\omega_n)$ field for $\omega_n = 0$. However,
the model  in Eq.~\eqref{eq:sxy}  is quantum mechanical, and the quantum fluctuations
are described by the finite Matsubara frequency (i.e.  $\omega_n \neq 0$)
components of $\theta_i(\omega_n)$. The latter and the classical (i.e. thermal) configurations
described by $\theta_i(\omega_n = 0)$ are coupled non-linearly  through the Josephson
coupling term $\propto J^0_{ij} \cos(\theta_i(\tau) - \theta_j(\tau))$. At low temperatures,
both quantum and classical fluctuations must be taken into account. This means that we must
obtain the effective classical limit of the quantum rotor model by integrating out the quantum
fluctuations described by the $\theta_i(\omega_n \neq 0)$ components of the phase. This is
especially  important for the anisotropic XY model because, as we drive the system
towards the extremely anisotropic limit where $t_y/t_x\ll 1$, the BKT transition temperature
$T_{c}$ tends to zero (cf. Figs.~\ref{Tc1} and \ref{phase}).

 In order to carry out the integration of the quantum fluctuations, we shall
 rely upon the self-consistent harmonic approximation (SCHA).~\cite{Cazalilla2,Feynman} Thus, we shall assume that,
 below $T_{c}$, the quantum rotor model of Eq.~\ref{eq:sxy} can be approximated by an anisotropic
 Gaussian model:
\be
S_{\mathrm G}[\theta]=\int^{\beta}_0d\tau\ \left[\sum_i\frac{(\partial_{\tau}\theta_i)^2}{2U}+\sum_{<i,j>}J_{ij}(\theta_i-\theta_j)^2\right]
\label{S_G}
\ee
where $J_{ij}$ is the  effective Josephson coupling  renormalized by the  interactions and the thermal fluctuations.
The derivation of a self-consistent equation for $J_{ij}$ is given  in Appendix \ref{a1}.

\subsection{Josephson Coupled Tomonaga-Luttinger Liquids and SCHA}

  For small values of the anisotropy ratio(i.e. for $t_y/t_x\to 0$), it is convenient to consider a different limit of the
the anisotropic Bose-Hubbard  model introduced in Eq.~\eqref{eq:bhm}. Indeed, for $t_{y} = 0$, Eq.~\eqref{eq:bhm}
reduces to an array of uncoupled 1D  Bose gases. For temperatures $T \ll t_x$, an interacting 1D Bose gas is
 known to behave as a Tomonaga-Luttinger liquid (TLL).~\cite{Cazalilla_RMP} Upon
 restoring a small $t_y( \ll t_x)$ coupling between the TLLs, the resulting system is an
 array of weakly coupled TLLs, which is described by the following effective Hamiltonian:~\cite{Cazalilla3,Cazalilla_RMP}
\begin{multline}
H_{\mathrm {CTLL}} = \frac{v}{2\pi} \sum_{i=1}^{L_y}  \int dx \left[ K \left( \partial_x \theta_i \right)^2 +
K^{-1} \left(\partial_x\phi_i \right)^2 \right] \\
- \frac{g^0_J v}{\pi a^2_0}\sum_{i=1}^{L_y} \int dx \, \cos\left[ \theta_i - \theta_{i+1} \right],\label{TLLs}
\end{multline}
where $v$ is the sound velocity, $K$ is the Luttinger parameter characterizing the decay of correlations, $a_0\approx a$ is short-range cutoff, and  $g^0_J \simeq 2 \pi t_y \rho_0 a^2_0/v$. The fields $\frac{1}{\pi}\partial_x\phi_i(x)$ and $\theta_i(x)$ describe the (long wavelength) density and  phase fluctuations of  the 1D interacting Bose gas at site $i=1, \ldots, L_y$ of the array.

 In order to obtain a phase-only description, we integrate out the density fields $\phi_i(x)$ in Eq.(\ref{TLLs}) and thus obtain
the following action for the array of weakly  coupled TLLs:
\begin{multline}
S_{CTLL}[\theta_i] = \frac{K}{2\pi}\sum_{i=1}^{L_y}\int^{\beta}_0d\tau\ \int^{L_x}_0d x \left[ \frac{(\partial_{\tau}\theta_i)^2}{v}+ v (\partial_{x}\theta_i)^2\right]\\
-\frac{g^0_J v}{\pi a^2_0}\sum_{i=1}^{L_y}\int^{\beta}_0d\tau\ \int^{L_x}_0d x \cos(\theta_i-\theta_{i+1})\label{eq:Sc1D},
\end{multline}
It is now possible to apply the SCHA to this model by  approximating the non-linear Josephson coupling in $S_{TLL}[\theta_i]$ by a Gaussian coupling:
\begin{multline}
S_{\mathrm{G}}[\theta_i] = \frac{K}{2\pi}\sum_{i=1}^{L_y}\int^{\beta}_0d\tau\ \int^{L_x}_0d x \left[ \frac{(\partial_{\tau}\theta_i)^2}{v}+ v (\partial_{x}\theta_i)^2\right]\\
+\frac{g_J v}{\pi a^2_0}\sum_{i=1}^{L_y}\int^{\beta}_0d\tau\ \int^{L_x}_0d x \, (\theta_i-\theta_{i+1})^2,\label{eq:TLL_G}
\end{multline}
where $g_J$ is the  effective SCHA  coupling. It can be computed by solving the equation in Appendix~\ref{a2}.

\subsection{BKT Transition and The sine-Gordon Model}\label{sec:sg}

The advantage of the Gaussian models (either~(\ref{S_G}) or (\ref{eq:TLL_G})), obtained after the application of the SCHA
approximation, is that they allow for readily integrating out the "quantum components" of the phase field (i.e. the
$\omega_n\neq 0$ components of $\theta$). We can thus obtain, in the continuum limit where the variation of the phase is slow
over the scale of the lattice parameter, a classical Gaussian model
%
\begin{equation}
S_{\mathrm{C}-G}[\theta] = \frac{1}{2} \int dx dy \left[ K_x \left( \partial_x \theta \right)^2 + K_y \left(\partial_y \theta \right)^2 \right],
\label{eq:cg}
\end{equation}
where the expressions for stiffnesses  $K_x$ and $K_y$ depend on the starting Gaussian model:
$K_x =  \beta J_x$ and $K_y =  \beta J_y$, for Eq.~\eqref{S_G}, and $K_x = \beta K v/(a\pi)$ and $K_y = \beta g_J/(a\pi)$,
for Eq.~\eqref{eq:TLL_G}. Interestingly enough, the role of the anisotropy in the continuum limit  description based on \eqref{eq:cg}
seems to be rather minor. This can be seen by rescaling the coordinates $x \to \eta^{1/2} x$ and $y\to \eta^{-1/2} y$, where
$\eta = \sqrt{K_x/ K_y}$, yielding the following isotropic Gaussian model:
\begin{equation}
S_{\mathrm{C}-G}[\theta] = \frac{K_{\beta}}{2} \int d\mathbf{r}\, \left( \nabla \theta \right)^2, \label{eq:cg2}
\end{equation}
where $K_{\beta} = \sqrt{K_x K_y}$. Note that, in a finite system,
the rescaling also affects the system dimensions: $L_x \to L_x \eta^{1/2}$
and $L_y \to L_y \eta^{-1/2}$. This observation will be important below.

 Eq.~(\ref{eq:cg2}) can be regarded as the \emph{na\"ive} continuum limit of Eq.\eqref{eq:cxy} and it
can only describe the (thermal) phase fluctuations within the superfluid phase of  Eq.~(\ref{eq:bhm}).
Thus, this model can only capture the algebraically decaying phase correlations characterizing the superfluid
phase of the XY model (cf. Sec.~\ref{sec:xy}). However, it is completely unable to capture the vortex and anti-vortex
unwinding that ultimately drives the BKT transition.

 In order to capture the possibility of topological excitations that ultimately lead to the BKT transition,
 we need to take a step back to the original XY model, either Eq.~(\ref{eq:sxy}) or Eq.~(\ref{eq:Sc1D}), and acknowledge that
by relying on  the SCHA, since we have   neglected the possibility of topological configurations of the phase where the
latter jumps by multiples of $2\pi$ from a given lattice site to a neighboring site. Thus, the right way to proceed would have  been to start from the quantum XY model
(or better, from the Bose-Hubbard model of Eq.~(\ref{eq:bhm})) and, after integrating out the quantum components of the
phase (and density) fields, to arrive at an effective classical XY model like Eq.~\eqref{eq:cxy}, with properly renormalized
parameters. The latter, via a  duality transformation,~\cite{Jose,Nagaosa} can be mapped onto the sine-Gordon model,
\be
S_{\mathrm sG}=\int d\bfr\: \left\{ \frac{\left[\nabla\phi(\bfr)\right]^2}{2 K^{(0)}_{\beta}}-\frac{2 g^{(0)}}{a^2} \cos2 \pi \phi (\bfr) \right\},
\label{S_sG}
\ee
where $\phi(\bfr)$ is a field that is dual~\cite{Jose,Nagaosa}  to $\theta(\mathbf{r})$ and $g^0 \propto  e^{-E_c/k_BT}$  is the so-called vortex fugacity with  $E_c$ being the vortex core energy. The classical 2DXY  and the sine-Gordon models belong to the same
universality class, which means that, near the BKT transition they provide an equally accurate description of the long-wave length
phenomena.  For the  2DXY universality class, Nelson and Kosterlitz
have shown~\cite{Nelson}  using the renormalization group (RG) that, at the critical temperature for the BKT transition, $T_c$,
the \emph{renormalized}  phase-stiffness ($K_\beta^{(R)}$) exhibits a universal jump:
\begin{align}
K^{(R)}_{\beta}(T\rightarrow T^{-}_c) &=\frac{2}{\pi}, \\
K^{(R)}_{\beta}(T\rightarrow T^{+}_c) &= 0. \label{eq:nelson2}
\end{align}
The renormalized stiffness $K^{(R)}_{\beta}$ satisfies a set of differential RG equations, which describe the `flow' of the sine-Gordon parameters (which correspond to $K^{(0)}_{\beta}$  and $g^{(0)}$ at the scale of the lattice parameter $a$)  as the system \emph{classical} degrees of freedom are coarse-grained in the vicinity of the BKT transition. Thus,  RG equations  determine the  long wavelength properties of the system, or, in other words, the phase of system: For $K^{(R)}_\beta >2/\pi$ (i.e for $T < T_c$), the coupling of the non-linear term  ($\propto \cos 2\pi \phi$) in Eq.~\eqref{S_sG}, which is responsible for the creation of vortex-anti-vortex  pairs, is renormalized down to zero, leading us back to the Gaussian model (cf. Eq.~\ref{eq:cg2}) that describes  the superfluid phase,
but with a renormalized value of the stiffness equal to $K^{(R)}_{\beta}$. On the other hand, when  $K^{(R)}_\beta<2/\pi$ (for $T > T_c$), the vortex-anti-vortex pairs unbind, which means that the coefficient of the $\cos 2\pi \phi$ term grows as the system is coarse grained. The unbinding disorders the system thus destroying the superfluidity (i.e. $K^{(R)}_{\beta}\to 0$), and thus the system becomes a normal Bose fluid.

However, it must be pointed out that the derivation of the sine-Gordon model from the original Bose-Hubbard model (cf. Eq.~\ref{eq:bhm})
or the  quantum XY model, Eq.~\eqref{eq:sxy} is very hard to carry out in practice. The reason is
that the integration of the $\omega_n \neq 0$ components of the phase cannot be performed exactly due to the non-linear nature of the Josephson coupling.
Thus, in this work we have chosen an alternative route,
which involves using the SCHA to obtain the Gaussian model with effective parameters, $K_x$ and $K_y$, from which we can
obtain an \emph{approximation} to the renormalized stiffness at $T_c$: $K^R_{\beta}(T_c) \approx K_{\beta}(T_c)  = \sqrt{K_x(T_c)
K_y(T_c)}$. As we shall show below by explicit comparison with QMC results, the SCHA provides a reasonably accurate estimate of the superfluid parameters even in an anisotropic Bose system where $T_c$ is driven
to zero. Within this framework, an approximation to the critical temperature is
determined from the condition that
\be
K_{\beta}(T_c) =\frac{2}{\pi}.
 \label{critical}
\ee
Note that, since $K_{\beta}$ is not the actual renormalized stiffness, it does not necessarily  vanish for $T > T_c$. However,
in accordance with \eqref{eq:nelson2} we  impose this fact by hand.

\subsection{QMC simulation on Bose-Hubbard model}\label{sec:qmc}

In order to validate the previously described approximations, we have carried \emph{ab initio} QMC simulations of the Bose-Hubbard model, Eq.(\ref{eq:bhm}), using the worm algorithm.~\cite{Pollet1} Earlier work~\cite{Barbara,Pollet} on isotropic 2D interacting Bose systems has shown that this algorithm can be used to study the KT transition.  However, as pointed out by Prokof'ev and Svistunov in
Ref.[~\oncite{prokofev_anisotropic}], the helicity modulus  depends strongly on the aspect ratio of the lattice employed in the QMC simulation, i.e. when the thermodynamic limit ($L_{x,y}\to\infty$) is taken by keeping $L_x/L_y$ fixed in  isotropic systems where
$t_x=t_y$. As a result, the definition of superfluidity and its transition temperature can be different for different aspect ratios. The reason
is that, as  $L_x/L_y$ is varied away from unity, the criticality of the system  also undergoes a crossover from a classical 2D XY  to 1D XY
universality class. In the latter case, $T_c$ and the helicity modulus vanish. The crossover would be complete when we able to
conduct simulations up to the thermodynamic limit. However, in finite systems finite-size effects prevent the system from completely
reaching the 1D XY fixed point.

  In this work, we focus on the effect of the hopping anisotropy, $\eta_0 = t_y/ t_y\neq 1$, on the superfluid properties. Therefore, we
 must first determine a physically sensible prescription to obtain the helicity modulus and hence the BKT transition temperature. To this end, in our QMC simulations we have chosen a value of the system aspect ratio, $L_x/L_y$, such that the excitation energy of a unit quantized flux is the same in both directions in the noninteracting limit, i.e. $t_x\left(\frac{2\pi}{L_x}\right)^2=t_y\left(\frac{2\pi}{L_y}\right)^2$ or $L_x/L_y=\sqrt{t_x/t_y}$.  For example, for $t_y/t_x=0.1$, we use $L_x=100$ and $L_y=32$ so that $L_x/L_y=3.125\simeq \sqrt{t_x/t_y}=\sqrt{10}=3.1622$.  The rationale for this choice is  explained in what follows.

 The helicity modulus  can be defined as:~\cite{Fisher,Pollock}
\begin{equation}
 \gamma_{x,y}=\frac{2 \Delta F(\phi_{x,y})}{\Omega(\phi_{x,y}/L_{x,y})^2},
\end{equation}
 where $\Omega = L_x L_y$ is the system area and   $\Delta F(\phi_{x,y})$ is the free energy change due to an
 infinitesimal phase twist $\phi_{x,y}$ applied at boundaries of the system. However,
in a QMC simulation,   the helicity modulus  can be also obtained  from
the winding number fluctuations $\langle W^2_{x,y}\rangle$:\cite{Pollock,prokofev_anisotropic}
\be
\gamma_{x, y}= T \frac{L_{x,y}}{L_{y,x}}\langle W_{x,y}^2\rangle,
\label{gammaQMC}
\ee
where $\langle W_{x}^2\rangle$ ($\langle W_{y}^2\rangle$) are the winding-number fluctuations along $x$ ($y$) direction.

 In continuum systems, the helicity
 modulus, $\gamma$,  can be related to a quantity with dimensions of density, namely the superfluid density $\rho_s$,
 by means of the equation:
 \begin{equation}
 \gamma = \frac{h^2}{m} \rho_s \label{eq:rhos}
 \end{equation}
 where $m$ is the particle mass. In lattice  systems, a natural generalization of \eqref{eq:rhos}
is obtained by replacing $m$ by the effective mass, $m^*_{x,y}$ ,which may be direction dependent.
Indeed, for free particles, $\frac{\hbar^2}{ m^*_{x,y}} \sim 2  t_{x,y}$, which implies that
 \begin{equation}
 \gamma_{x,y} = 2 t_{x,y}  \rho_s.
 \end{equation}
  Therefore, the choice of aspect ratio $L_x/L_y=\sqrt{t_x/t_y}$ means  (cf. Eq.~\ref{gammaQMC}) that
  $\langle W_{x}^2\rangle=\langle W_{y}^2\rangle$ and thus the superfluid density $\rho_s$
 alone determines the helicity modulus in both directions.

\begin{figure}[]
\includegraphics[width=8.5cm]{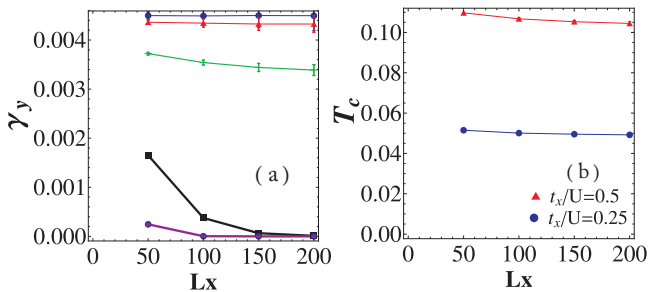}
\caption{ (a) shows the helicity modulus in the $y$ direction, ${\gamma_y}$ obtained by QMC as a function
of the longest side, $L_x.$ We use $\eta_0= J_y^0/J_x^0=0.02$ , $t_x/U= 0.25$ , and $\rho_0\simeq 0.8$. From top to bottom lines are for $T/U=0.01$ to $0.09$ in steps of $0.02$. (b) $T_c$ as a function
of $L_x$ for $\eta_0= J_y^0/J_x^0=0.02.$ $t_x/U=0.5,$ $\rho_0\simeq 0.7$ for red triangles and $t_x/U=0.25,$ $\rho_0\simeq 0.8$ for blue dots. }
\label{finite size}
\end{figure}

 Next, let us  assess the importance of finite size effects using the above choice for the system aspect ratio.
 In Fig. \ref{finite size}(a), we show the helicity modulus in the $y$ direction, ${\gamma_y}$, as a function
of the longest side, $L_x$, where  $t_x/U=0.25$, $\rho_0\simeq 0.8$ and $\eta_0= J_y^0/J_x^0=0.02.$ By the variation of $T/U$ from $0.01$ to $0.09$, we find that, for $L_x\geq100$ and $T\leq0.05$, ${\gamma_y}$ is almost unchanged. In Fig. \ref{finite size}(b),
we show $T_c$ as a function of $L_x$ for $\eta_0= J_y^0/J_x^0=0.02$. It can be seen that the variation of $T_c$ (see Sec.~\ref{KTtransition} for an explanation of how $T_c$ is estimated from the QMC data) with $L_x$ is less than $5\%$. These results justify that neglecting
finite-size effects  on the helicity modulus and $T_c$  for the typical system sizes employed in our QMC simulations ($L_x>100$).

\section{Small to intermediate anisotropy}\label{sec:small_anisotropy}

\subsection{Inside the Superfluid Phase}\label{sec:large_eta}

 We first discuss the results of SCHA for the XY model, which approximates Eq.~\eqref{eq:sxy} by the Gaussian model of Eq.~\eqref{S_G} with an effective quadratic coupling, $J_{ij}(T,U,n_0)$. The derivation of the equation for  $J_{ij}$ is given in Appendix \ref{a1} (cf. Eq.~(\ref{renormalized_J})).
%
%
%
We note that the non-linear Josephson term in Eq.(\ref{eq:sxy}) couples  all Matsubara frequencies, and therefore,
in SCHA, the renormalized $J_{ij}$ in Eq.(\ref{S_G}) acquires a temperature dependence.

\begin{figure}[]
\includegraphics[width=8.5cm]{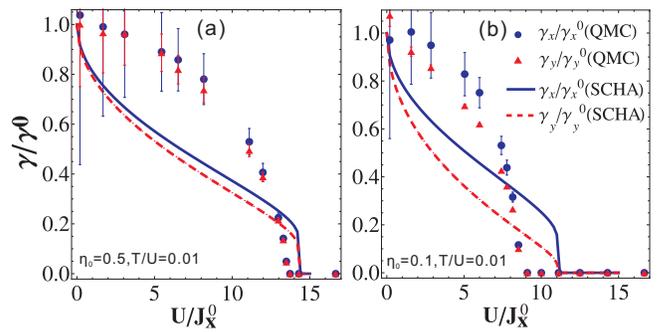}
\caption{The renormalized helicity modulus, ${\gamma_\alpha}/{\gamma_\alpha^0}$, as a function
of the interaction strength, $U/J^0_x,$. Note that, on both figures,  we keep constant the values of $U$, $T$, and $\eta_0= J_y^0/J_x^0=0.5$ and $0.1$, and change $J_{x}^0$. Blue solid (red dashed) lines represent the results in the $\alpha=x(y)$ directions. For comparison, we also show the numerical results obtained by QMC in filled circles and triangles together. See the text for more details of comparison.}
\label{renormalization_SF}
\end{figure}

 From the continuum limit of the Gaussian model obtained from the SCHA (cf. Eq. \ref{S_G}),
the helicity modulus can be read off: $\gamma_{x,y}= 2 J_{x,y}$.
Hence, we can also define anisotropy ratio as $\eta=\gamma_y/\gamma_x = J_y/J_x$.  For later purposes,
it is also worth introducing the \emph{bare} (i.e. unrenormalized) system parameters:
$\gamma^0_{x,y}= 2 J^0_{x,y}  =2   t_{x,y} \rho_0$ ($\rho_0$ is the mean lattice occupation and $a$ the lattice parameter)
and the bare anisotropy ratio $\eta_0=\gamma^0_y/\gamma^0_x=J^0_y/J^0_x=t_y/t_x$. In what follows, we
compare the results of $\gamma_{x,y}$ obtained from SCHA and QMC within anisotropic superfluid (SF) phase.

In Fig. \ref{renormalization_SF} we show the ratio of the renormalized to the bare helicity moduli, $\gamma_{\alpha}/\gamma^0_{\alpha}$ as a function of the interaction strength, $U/J_x^0$. The bare anisotropy ratio parameter is chosen to be $\eta_0=0.5$ ( Fig. \ref{renormalization_SF}(a)) and $\eta_0=0.1$ (Fig.~ Fig. \ref{renormalization_SF}(b)). Here we keep both $U$ and $T$ constant but change $J_x^0$ and $J_y^0$ in order to comparison with QMC data more easily. As expected, increasing the strength of interactions, that is,
increasing $U/J_x^0$, suppresses superfluidity as that both $\gamma_{x}$ and $\gamma_y$ decrease.
Note that, within the SCHA, even a weak interaction can have a strong effect  on the renormalized
helicity modulus, $\gamma_{x,y}$. Indeed, when the interaction is larger than a critical value, the helicity modulus drops to zero in both directions discontinuously, and the system becomes a normal fluid without phase stiffness. This is a feature of the SCHA, which
wrongly predicts the interaction-driven transition between the SF and the Normal fluid (which at $T = 0$ corresponds
to the SF to Mott insulator quantum phase transition) to be of first order.


In the same figure, we also show numerical results of our QMC simulation for comparison. We can see that, although the ratio of the renormalized to the bare helicity modulus obtained from QMC exhibits qualitatively the same behavior as the SCHA,  it does not show strong renormalization effects predicted by the SCHA at small $U/J_x^0$. Furthermore, at larger $U/J_x$, both $\gamma_x/\gamma^0_x$ and $\gamma_y/\gamma^0_y$ vanish rather smoothly.

\begin{figure}[]
\centering
\includegraphics[width=8.5cm]{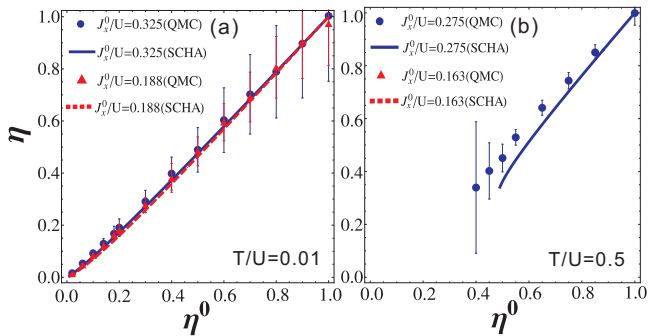}
\caption{Ratio of the renormalized helicity moduli, $\eta$ vs.
its bare (unrenormalized) value, $\eta_0$, for (a) $T/U=0.01$ and (b) $T/U=0.5$.  We choose different values of $J^0_x/U$: (a) $J^0_x/U=0.325$ and $J^0_x/U=0.188$ and  (b) $J^0_x/U=0.275$ and $J^0_x/U=0.163$, and  vary $\eta_0$. Note that
in (b), for $\eta_0 \lesssim0.5$, the temperature is higher than the BKT transition temperature and therefore both $\gamma_x$ and
$\gamma_y$ vanish.  We also show the results of our QMC simulations for comparison.
 Both QMC and SCHA yield results in excellent agreement, suggesting that the
anisotropy ratio of the helicity modus is barely renormalized by interaction and finite-temperature effects. This is consistent
with the SF phase being described by an isotropic Gaussian field theory, as discussed in Sec.~\ref{sec:sg}.}
\label{eta_U}
\end{figure}

In order to better understand how  finite temperature  and interactions influence the anisotropy ratio of the helicity modulus,
we show in Fig.~\ref{eta_U}(a) and (b) the renormalized helicity ratio, $\eta=\gamma_y/\gamma_x$ vs. the bare one $\eta_0= \gamma^0_y/\gamma^0_x=t_y/t_x$. Results obtained both from the SCHA and QMC are shown together for comparison. We
see that, at low temperatures ($T/U=0.01$,  Fig.~\ref{eta_U}(a)), when the system is  deep in the superfluid phases, the anisotropy
is barely renormalized, i.e. $\eta \simeq \eta_0$ and indeed our QMC results agree well with the SCHA predictions for $\eta$. Interestingly, this result holds also true at much higher temperatures (cf. Fig.~ \ref{eta_U}(b)), except for the fact that, for small values of $\eta_0$  the system becomes a normal gas (i.e. the temperature used in the simulation $T/U  = 0.5$  is larger than the BKT transition
temperature,  $T_c$, for these highly anisotropic systems). This is because as $T_c$ of an anisotropic superfluid ($\eta_0<1$) becomes smaller, the renormalization of the helicity ratio also becomes more significant near the phase transition boundary. The agreement between SCHA and QMC results is very good.

  Thus,  we find that, although the SCHA  and QMC  yield  different values for renormalized  helicity moduli,
 $\gamma_x$ and $\gamma_y$,  QMC shows that the renormalized anisotropy ratio
 $\eta = \gamma_y/\gamma_x$ is  barely affected by interaction and/or temperature effects.
 This is consistent with the SF phase being described,
 in the continuum limit, by an isotropic Gaussian field theory (cf. Eq.~\ref{eq:cg2} in  Sec.~\ref{sec:sg}), which is
also correctly captured by the SCHA.

\begin{figure}[htbp!]
\centering
\includegraphics[width=8.5cm]{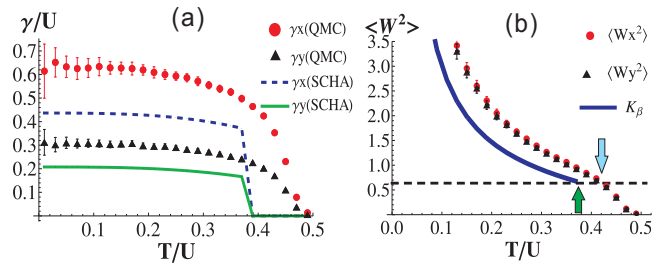}
\caption{The helicity modulus as a function of temperature for a fixed interaction $U$. The bare tunneling are $t_x/U=0.5$, $t_y/U=0.25,$ and the density is $\rho_0\simeq 0.63$. (a) shows results including interaction renormalization within SCHA, compared with the QMC results in dots and in triangulars. (b) shows the winding number fluctuation in QMC, and $K_\beta$ obtained analytically from the SCHA to XY model, as a function of temperature. The horizontal dashed line indicate the universal number, $2/\pi$. The intersection of the curves (dots) and the horizontal lines gives the $T_c$ in SCHA (QMC), marked by arrows.}
\label{bosonSFbyRG}
\end{figure}
\subsection{Near the BKT transition}\label{KTtransition}

In Fig. \ref{bosonSFbyRG},  we show the helicity modulus (proportional to superfluid density, cf. Eq.~\ref{eq:rhos}) as a function of temperature. 
The bare single particle tunneling amplitude is $t_x/U=0.5$ and $t_y/U=0.25$ respectively, and the filling fraction
is $\rho_0\simeq 0.63$.  In Fig. \ref{bosonSFbyRG}(a), the results of the helicity moduli obtained from the SCHA and
QMC are compared. We thus see that, compared to  the QMC results, the SCHA overestimates the temperature dependence
of the helicity modulus in both directions roughly by a factor of one point five.


 In Fig. \ref{bosonSFbyRG}(b), we show how the BKT transition temperature $T_c$ is determined from both the analytical results
 of SCHA and the QMC data.  In the case of the SCHA, we compute the phase stiffness as discussed in Sec.~\ref{sec:sg}, i.e.
 from  $K_{\beta} = \beta \sqrt{J_x J_y}$, where $J_x$ and $J_y$ are solutions to the SCHA equations for given $T$ and $U, J^0_x, J^0_x$ values. Hence, $T_c$ is found by  varying the temperature until $K_{\beta}  = \frac{2}{\pi}$  (cf. Sec.~\ref{sec:sg}).

  As to the QMC data,  $T_c$ is obtained as follows:  As anticipated in Sec.~\ref{sec:qmc},
by choosing  $L_x/L_y=\sqrt{t_x/t_y}$ we find that the winding number fluctuations (red dots and black triangles in Fig.~\ref{bosonSFbyRG}(b)) in the $x$ and $y$ directions essentially coincide. Furthermore,  the $\langle W^2_{x,y}\rangle$ show  kink at a temperature, which is essentially equal to the one obtained by  requiring that $T_c$ (cf. Sec.\ref{sec:sg}):
\begin{align}
K^{QMC}_\beta(T_c)&= \frac{1}{T_c}\sqrt{\gamma_x(T_c)\gamma_y(T_c)}\notag\\
&=\sqrt{\langle W_{x}^2(T_c)\rangle\langle W_{y}^2(T_c)\rangle}\simeq \langle W^2_{x,y} \rangle =\frac{2}{\pi}
\end{align}
where  Eqs. \eqref{critical} and \eqref{gammaQMC} have been used.  In Fig. \ref{bosonSFbyRG}(b), we have indicated the universal
value of $\frac{2}{\pi}$ by a horizontal line. As explained in Sec.~\ref{sec:sg}, in the SCHA, we assume that $K_{\beta}$ vanishes for
for $T > T_c$. However, in the QMC calculations, finite-size effects round off the expected thermodynamic-limit discontinuity of $K^{QMC}_{\beta}$  at $T = T_c$. Yet, as discussed in Sec.~\ref{sec:qmc}, the value of $T_c$ estimated from the kink in the
Monte Carlo data is converged for system sizes that we used ($L_x > 100$). Finally,  the comparison of $T_c$ as obtained from QMC
and SCHA is shown in Fig. \ref{Tc} and  will be explained in more detail further below.


\section{Large anisotropy} \label{sec:large_anisotropy}
\subsection{SCHA for the Coupled TLLs}
\begin{figure}[]
\includegraphics[width=8.5cm]{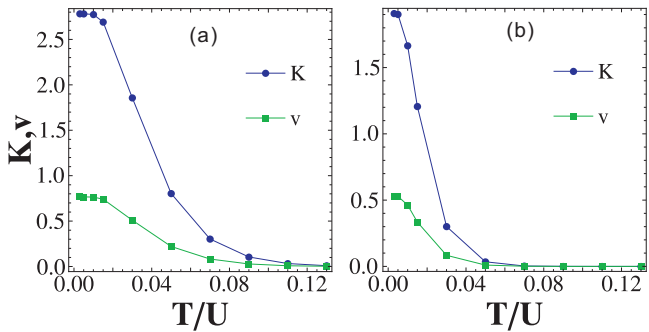}
\caption{The Luttinger parameters $K$ and $v$ as a function
of the temperature, $T/U.$ For the 1D Bose-Hubbard model, we fix $L_x=150$ for both plots, and (a) $t_x/U = 0.5, t_y/U=0, \rho_0\simeq 0.7$ and (b) $t_x/U = 0.25, t_y/U=0, \rho_0\simeq 0.8$ respectively. }
\label{KvsLT}
\end{figure}
To begin with, let us note that, for the 1D Bose-Hubbard model,  the Luttinger parameters $K$ and $v$ that determine the properties of the TLLs in the decoupled limit (cf. Eq.~\ref{TLLs}
for $g^0_J = 0$)  cannot be analytically obtained  for general lattice fillings and values of $U/t_x$
(Eq.~\ref{eq:bhm} for $t_y = 0$). Thus, in order to extract the  Luttinger liquid parameter,
$K$, and sound velocity, $v$,   we have carried out additional QMC calculations for the 1D Bose-Hubbard model to extract these parameters. Using the relations
$v/K=1/\pi \kappa$ and $v K = \pi L_x T \langle W_{x}^2 \rangle$,   where $\kappa={\partial \rho}/{\partial\mu}$ is the compressibility and
$\langle W_{x}^2 \rangle$ is the  winding number fluctuation along the $x$ direction for $T/U\ll 1$.~\cite{Cazalilla5}  In Fig. \ref{KvsLT} the numerical Luttinger parameters $K$ and $v$ as a function
of the temperature are shown, $T/U$, for a large large size of the 1D system of $L_x=150$. The parameters
characterizing one (decoupled) TLL correspond to the extrapolation of this results to very low temperature.
Thus, for $T/U=0.005$ we find $K \simeq 2.77$ and $v \simeq 0.77$ for $t_x/U=0.5, \rho_0\simeq 0.7$ in Fig. \ref{KvsLT}(a), and  $K \sim1.91,$ $v \simeq 0.53$ for $t_x/U=0.25, \rho_0\simeq 0.8$ in Fig. \ref{KvsLT}(b).

\begin{figure}[htbp!]
\centering
\includegraphics[width=8.3cm]{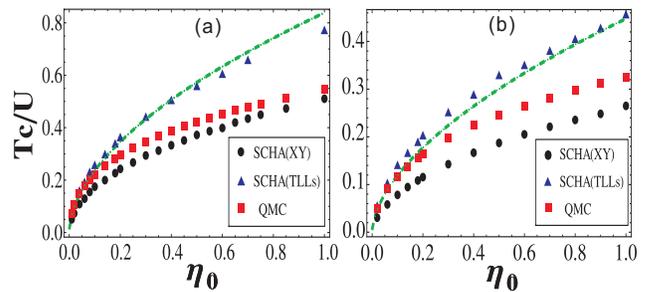}
\caption{$T_c$ as a function of the bare anisotropy ratio, $\eta_0= J^0_y/J^0_x = t_y/t_x$ for $t_x/U=0.5$, $\rho_0\simeq 0.65\pm0.05$ ($J^0_x \simeq 0.325$)  (a) and $t_x/U=0.25$, $\rho_0\simeq 0.75\pm0.05$ ($J^0_x/U\simeq 0.1825$)  (b)
The blue triangles are the results obtained by SCHA to the XY model. The black dots correspond to the results obtained by SCHA to an array of coupled TLLs. The Red squares are the QMC data.   $T_c$ is determined by the methods discussed in Section \ref{KTtransition}. On both panels, the green dash curve is a  fit to the scaling behavior of Tc with the bare anisotropy ratio $\eta_0$ yielding $T_c/U \simeq 0.837 \eta_0^{0.55}$  (a) $T_c/U \simeq 0.448 \eta_0^{0.575}$  (b). }
\label{Tc}
\end{figure}

Next, we describe the result of applying the SCHA to the system of coupled TLLs. Compared to the case of
small anisotropy discussed above (Eq.~\ref{renormalized_J}), in this case only $J_y$ is renormalized, and
all the interaction dependence of $J_y$ enters through the  Luttinger parameters. However,  as discussed in sec.~\ref{sec:sg},
the system of coupled TLLs at finite temperature also belongs to the 2DXY universality class~\cite{Cazalilla2,Cazalilla3}.
Thus, the BKT critical temperature can be found from the equation:
\be
K_{\beta} = 2 \frac{\sqrt{K v J_y(T_c)/2\pi}}{T_c}=\frac{2}{\pi},\label{tctll}
\ee
In  the SCHA calculations, we have chosen the short-distance cut-off such that $K v a_0/2\pi\simeq J_x$, when comparing the TLL-Gaussian model of Eq.~\eqref{eq:TLL_G} with the Gaussian model in Eq.~\eqref{S_G}.

In Fig. \ref{Tc} (black dots), we show the BKT critical temperature, $T_c$ computed using the SCHA, and QMC as a function of the bare anisotropy ratio.
As discussed above, $T_c$  goes to zero gradually as the bare anisotropy ratio $\eta_0$ becomes larger, reflecting the fact that there is no superfluid phase transition at finite temperature in 1D system. From both panels in Fig. \ref{Tc}, it can be seen that the SCHA to  XY model provides a reasonably good description of $T_c$ (compared to the QMC results) for $\eta_0\approx 1$ and  weak interactions (Fig.~\ref{Tc}(a)), but it deviates from the QMC results for stronger interactions (Fig.~\ref{Tc}(b)) and small $\eta_0$. On the other hand,  the results obtained by applying the SCHA to an array of coupled TLLs are found to be closer to the QMC results for $T_c$ in the large anisotropy regime (i.e. small $\eta_0$). These results are consistent with the expectation that the SCHA to the XY model  should be more accurate in the small anisotropy regime, whereas applying the SCHA to an array of coupled TLLs becomes
a better approximation in the limit of large anisotropy.

\subsection{RG scaling for critical temperature}

Besides of the numerical calculations of BKT critical temperature, from our QMC data  we can also extract the scaling behavior of Tc with anisotropy ratio $\eta_0$. This can be compared with the results obtained by the renormalization group flow of the Josephson coupling in Eq \eqref{eq:Sc1D}, which described by the differential equation:~\cite{Cazalilla2,Cazalilla3}
\begin{equation}
\frac{d g_J}{ d\ell} = \left(2 - \frac{1}{2K} \right) g_J. \label{eq:rg}
\end{equation}
where the flow parameter $ \ell = \ln \left( {a(\ell)}/{a_0}\right) = \ln \left({\Lambda(0)}/{\Lambda(\ell)} \right),$ with $a(\ell)=a_0 e^{\ell}.$
Since $K \in [1, +\infty)$ for the Bose-Hubbard
model is far from the critical point $K^* = {1}/{4},$  we can neglect the renormalization of $K$ and treat it as a constant.~\cite{Cazalilla2,Cazalilla3} Therefore,
the solution to  \eqref{eq:rg} reads $g_J(\ell) \simeq g_J(0) e^{(2-\frac{1}{2K})\ell}$. To complete the solution,
we need to recall that the bare (energy) cut-off $\Lambda(0)
\approx \: t_x, $ and $g_J(0) \simeq 2 \pi t_y \rho_0 a^2_0/v$. In order to estimate of the critical temperature at which the
system will enter the SF phase, we note that, at finite temperatures,  the RG flow is cut off  at the scale
$\Lambda(\ell) \simeq T$, and $g_J(T= T_c) \sim 1$. Hence, provided \eqref{eq:rg} provides accurate description of the e
entire flow (i.e. for small enough $g_J(0)$),  we have
\begin{equation}
T_c \simeq  C   {\eta_0 }^{\frac{1}{2- 1/2K}} = C {\eta_0 }^{\frac{2K}{4 K -1}}\label{eq:tcrg}
\end{equation}
where $C$ is a prefactor that depends on microscopic details of the model, and can be obtained by fitting above scaling law to the QMC results.  It is worth noting that~\cite{Cazalilla3}  the same scaling  law for  $T_c/t_x$ with $\eta_0$ can be also obtained using mean-field theory, i.e. by  assuming that $\langle e^{i \theta_n(0) } \rangle = \phi_0(T).$ However, as discussed in the Introduction, strictly speaking mean-field theory is inapplicable  in two dimensions due to the lack of BEC at finite temperatures.

In Fig. \ref{Tc} we use the values of the Luttinger parameters obtained earlier from QMC simulations of the 1D Bose-Hubbard model ($K \simeq 2.77$ and $v \simeq 0.77$ in Fig.~\ref{Tc}(a) and $K \sim1.91,$ and $v \simeq 0.53$ in Fig.~\ref{Tc}(b)) to fit  the scaling of $T_c$. In particular, the value of $K$ completely determines the exponent of the scaling law (cf Eq.\ref{eq:tcrg}), and thus, the only free parameter is the prefactor $C$. The  fit yields $T_c/U \simeq 0.837 \eta_0^{0.55}$  for the data on  Fig. \ref{Tc}(a) and
$T_c/U \simeq 0.448 \eta_0^{0.575}$ for the data on  Fig. \ref{Tc}(b). Using the TLLs parameters obtained from $T/U=0.005,$ the predicted $T_c$ of SCHA to TLLs are close to QMC calculations at large anisotropy regimes.

\section{conclusion} \label{sec:conclusion}

In summary, two different approaches, the SCHA and QMC, reveal the highly
nontrivial features of the helicity modulus and the BKT phase transition in the 2D Bose-Hubbard model with anisotropic hopping.  These characteristic features simulated by QMC using a specific system
aspect ratio, $L_x/L_y=\sqrt{t_x/t_y},$ is consistent with the rescaling of the effective sine-Gordon model. We show how the interaction and finite temperature effect influence the helicity modulus and find profound agreement of anisotropy of the helicity modulus given by the SCHA and QMC. As we drive the system towards the extremely anisotropic limit, the BKT transition temperature approaches to the absolute zero and the transition thus becomes a 3DXY quantum critical point at the end of a line of classical 2DXY critical points. In particular, through the RG scheme for the coupled TLLs, we obtain the scaling relation of Tc with anisotropy ratio. Employing ultra-cold atoms in an controllable optical lattice opens avenues to identify our results of 2D anisotropic Bose-Hubbard model.

\section{Acknowledgement}

This work is supported by NSC grants and NCTS at the same time.
MAC gratefully acknowledges the hospitality of NCTS (Taiwan) and
the financial support of the Spanish MEC through grant FIS2010-19609-C02-02.

\appendix


\section{Self-consistent Harmonic Approximation }\label{a1}

To find the optimally quadratic approximation to the XY-model,  we employ the self-consistent
Harmonic approximation (SCHA). In this approach,  the action of XY or quantum rotor model, Eq.~\eqref{eq:qrotor},
is approximated by  an anisotropic Gaussian model:
\be
S_{\mathrm{G}}[\theta]&=&\int^{\beta}_0d\tau\ \{\sum_i\frac{(\partial_{\tau}\theta_i)^2}{2U}+\sum_{<i,j>}J_{ij}(\theta_i-\theta_j)^2\}\nonumber\\
&=&\frac{1}{2}\sum_{\bfk,\omega_n} G_v^{-1}(\bfk,\omega_n)|\theta(\bfk,\omega_n)|^2,
\ee
where $\omega_n=2\pi T n$, and the single particle Green's function is given by
\be
 G_v^{-1}(\bfk,\omega_n)&=&\frac{\omega_n^2}{U}+\sum_{\alpha}8J_{\alpha}\sin^2 (k_{\alpha}\hat{a}_{\alpha}/2).
\ee
Next, we make use of Feynman's variational principle, which states that:
\be
F=-\frac{1}{\beta}\ln Z\leq \tilde{F}[G_v]=F_v+\frac{1}{\beta}\langle S[\theta]-S_{\mathrm{G}}[\theta]\rangle_v,
\ee
where $\langle\rangle_v$ denotes the average with respect to $S_{\mathrm{G}}[\theta]$ and $S$ is the
XY model action. Since
\be
e^{-\beta F_v}=\int D\theta e^{-S_v[\theta]}=\prod_{\bfk,\omega_n} G_v(\bfk,\omega_n)^{-1/2},
\ee
the first term of $\tilde{F}[G_v]$ is:
\be
F_v=-\frac{1}{2\beta}\sum_{\bfk,\omega_n} \ln G_v(\bfk,\omega_n).
\ee
The remaining contributions to $\tilde{F}[G_v]$ are
\be
&&\langle S_{\mathrm{xy}}[\theta]-S_{\mathrm{G}}[\theta]\rangle_v\nonumber\\
&=&\langle\int^{\beta}_0d\tau\ \{\sum_i\frac{1}{2U}(\partial_{\tau}\theta_i)^2-\sum_{<i,j>}2J^0_{ij}\cos(\theta_i-\theta_j)\}\rangle_v\nonumber\\&&-\langle S_{\mathrm{G}}[\theta]\rangle_v\nonumber\\
&=&\sum_{\bfk,\omega_n}\frac{\omega_n^2}{2U}G_v(\bfk,\omega_n)+\langle S_{\cos}\rangle_v- \mathrm{const.}
\ee
with
\be
\langle S_{\cos}\rangle_v=-\int^{\beta}_0d\tau \sum_{<i,j>}2J^0_{ij} \langle \cos(\theta_i-\theta_j)\rangle_v.\nonumber
\ee
Hence,
\be
 \langle \cos(\theta_i-\theta_j)\rangle_v&=&\textmd{Re}\left[e^{-\frac{1}{2}\langle (\theta_i-\theta_j)^2\rangle_v}\right] \nonumber\\
 &=&\textmd{Re}\left[^{ G_v(r_i-r_j,0)-G_v(0,0)}\right],
 \ee
by the cumulant expansion. Here
\be
G_v(r,\tau)=\frac{1}{\beta {\Omega}}\sum_{\bfk,\omega_n} e^{\textmd{i}\bfk r}e^{-\textmd{i}\omega_n \tau}G_v(\omega_n, \bfk)\nonumber,
\ee
is the single particle Green's function in real space.
Therefore we have,
\be
\langle S_{\cos}\rangle_v&=&-\int^{\beta}_0d\tau \sum_{<i,j>}2J^0_{ij}\textmd{Re}\left[e^{ G_v(r_i-r_j,0)-G_v(0,0)}\right]\nonumber\\
&=&-\beta \sum_i\sum_{\mathbf{t}=\hat{a}_x,\hat{a}_y}2 J^0_{\mathbf{t}}  \textmd{Re}\left[e^{ G_v(\mathbf{t},0)-G_v(0,0)}\right]\nonumber\\
&=&-\beta {\Omega}\sum_{\alpha}2 J^0_{\alpha}  \textmd{Re} \left[e^{ \frac{1}{{\Omega}\beta}\sum_{\mathbf{k},\omega_n}(e^{\textmd{i} {\mathbf{k}} \cdot \hat{a}_{\alpha}}-1)G_v(\mathbf{k},\omega_n)}\right].\nonumber\\
\ee
Upon combining above results and  finding the extrema of $\tilde{F}[G_v]$, i.e.
\be
\frac{\delta F'[G_v]}{\delta G_v(\bfk,\omega_n)}=0,
\ee
we find
\be
&&\frac{1}{G_v(\bfq,\omega_n)}\nonumber\\
&=&\frac{\omega_n^2}{U}+8 \sum_{\alpha}J^0_{\alpha}\sin^2({ q_{\alpha} \hat{a}_{\alpha}/2})e^{ \frac{1}{{\Omega}\beta}\sum_{\bfk,\omega_n}(e^{\textmd{i} \bfk \cdot \hat{a}_{\alpha}}-1)G_v(\bfk,\omega_n)}\nonumber\\
&\equiv&\frac{\omega_n^2}{U}+8\sum_{\alpha}J_{\alpha}\sin^2 (q_{\alpha}\hat{a}_{\alpha}/2).
\ee
Using the Matsubara sum
\be
\frac{1}{\beta}\sum_{\omega_n}G_v(\bfk,\omega_n)&=&\frac{U}{\beta}\sum_{\omega_n}\frac{1}{\omega_n^2+\omega_{\bfk}^2}\nonumber\\
&=&\frac{U}{2{\omega_{\bfk}}}\coth(\frac{\beta\omega_{\bfk}}{2}),
\ee
we conclude that
\be
\ln\frac{J_{\alpha}}{J^0_{\alpha}}&=&\frac{1}{\Omega}\sum_{\bfk}(e^{\textmd{i} \bfk \cdot \hat{a}_{\alpha}}-1)\frac{U}{2{\omega_{\bfk}}}\coth(\frac{\beta\omega_{\bfk}}{2}).\label{renormalized_J}
\ee
Note that $\omega_{\bfk}= 2\sqrt{2 U }\sqrt{J_x\sin^2 (k_x{a}/2)+J_y\sin ^2 (k_y{a}/2)}$ is the phonon (Bogoliubov) excitation energy.

\section{SCHA for Coupled TLLs}\label{a2}
Applying the methods of  previous section to the action of  Eq.~(\ref{eq:Sc1D}),
the following equation for the renormalized parameter $J_y(=g_J v/\pi a^2_0)$  is obtained:
\be
\ln\frac{g_{J}}{g^0_{J}}&=&\frac{v K^{-1}}{L_x L_y}\sum_{\bfk}\frac{e^{\textmd{i} \bfk \cdot \hat{{\bf y}}}-1}{\omega_{\bfk}}
\coth\left(\frac{\beta\omega_{\bfk}}{2}\right),
\ee
where
$\omega_{\bfk}= 2\sqrt{v^2( k_x/2)^2+\frac{2 \pi v J_y}{K}\sin ^2 (k_y/2)}.$

\end{document}